\documentclass[]{aa}
\usepackage{graphicx}
\usepackage{txfonts}
\usepackage{color}
\usepackage[colorlinks,citecolor=blue]{hyperref}
\usepackage{multirow}  
\newcommand{\angstrom}{\mbox{\normalfont\AA}}

\begin{document}

\title{Anatomy of the AGN in NGC\,5548}
\subtitle{IX. Photoionized emission features in the soft X-ray spectra}

\author{Junjie Mao\inst{\ref{inst_sron},\ref{inst_leiden}}
\and J. S. Kaastra\inst{\ref{inst_sron},\ref{inst_leiden}}
\and M. Mehdipour\inst{\ref{inst_sron}}
\and Liyi Gu\inst{\ref{inst_sron},\ref{inst_riken}}
\and E. Costantini\inst{\ref{inst_sron}}
\and G. A. Kriss\inst{\ref{inst_stsi}}
\and S. Bianchi\inst{\ref{inst_ur3}}
\and \\ G. Branduardi-Raymont\inst{\ref{inst_mscl}}
\and E. Behar\inst{\ref{inst_tiit}}
\and L. Di Gesu\inst{\ref{inst_ug}}
\and G. Ponti\inst{\ref{inst_mpe}}
\and P.-O. Petrucci\inst{\ref{inst_uga},\ref{inst_cnrs}}
\and J. Ebrero\inst{\ref{inst_esa}}
}

\institute{SRON Netherlands Institute for Space Research, Sorbonnelaan 2, 3584 CA Utrecht, the Netherlands\label{inst_sron} \\ \email{J.Mao@sron.nl}
\and Leiden Observatory, Leiden University, Niels Bohrweg 2, 2300 RA Leiden, the Netherlands\label{inst_leiden}
\and RIKEN Nishina Center, 2-1 Hirosawa, Wako, Saitama 351-0198, Japan\label{inst_riken}
\and Space Telescope Science Institute, 3700 San Martin Drive, Baltimore, MD 21218, USA\label{inst_stsi}
\and Dipartimento di Matematica e Fisica, Universit\`{a} degli Studi Roma Tre, via della Vasca Navale 84, 00146 Roma, Italy\label{inst_ur3}
\and Mullard Space Science Laboratory, University College London, Holmbury St. Mary, Dorking, Surrey, RH5 6NT, UK\label{inst_mscl}
\and Department of Physics, Technion-Israel Institute of Technology, 32000 Haifa, Israel\label{inst_tiit}
\and Department of Astronomy, University of Geneva, 16 Ch. d’Ecogia, 1290 Versoix, Switzerland\label{inst_ug}
\and Univ. Grenoble Alpes, IPAG, F-38000 Grenoble, France\label{inst_uga}
\and CNRS, IPAG, F-38000 Grenoble, France\label{inst_cnrs}
\and European Space Astronomy Centre, P.O. Box 78, E-28691 Villanueva de la Ca\~{n}ada, Madrid, Spain\label{inst_esa}
\and Max Planck Institute fur Extraterrestriche Physik, 85748, Garching, Germany\label{inst_mpe}
}

\date{Received date / Accepted date}

 
\abstract
{The X-ray narrow emission line region (NELR) of the archetypal  Seyfert 1 galaxy NGC\,5548 has been interpreted as a single-phase photoionized plasma that is absorbed by some of the warm absorber components. This scenario requires those overlaying warm absorber components to have larger distance (to the central engine) than the X-ray NELR, which is not fully consistent with the distance estimates found in the literature. Therefore, we reanalyze the high-resolution spectra obtained in 2013--2014 with the Reflection Grating Spectrometer (RGS) aboard \textit{XMM}-Newton to provide an alternative interpretation of the X-ray narrow emission features. We find that the X-ray narrow emission features in NGC\,5548 can be described by a two-phase photoionized plasma with different ionization parameters ($\log \xi=1.3$ and $0.1$) and kinematics ($v_{\rm out}=-50$ and $-400~{\rm km~s^{-1}}$), and no further absorption by the warm absorber components. The X-ray and optical NELR might be the same multi-phase photoionized plasma. Both X-ray and optical NELR have comparable distances, asymmetric line profiles, and the underlying photoionized plasma is turbulent and compact in size. The X-ray NELR is not the counterpart of the UV/X-ray absorber outside the line of sight because their distances and kinematics are not consistent. In addition, X-ray broad emission features that we find in the spectrum can be accounted for by a third photoionized emission component. The RGS spectrum obtained in 2016 is analyzed as well, where the luminosity of most prominent emission lines (the \ion{O}{vii} forbidden line and \ion{O}{viii} Ly$\alpha$ line) are the same (at a 1 $\sigma$ confidence level) as in 2013--2014.}


\maketitle

\section{Introduction}
\label{sct:intro}
Two types of emission lines are commonly observed in the optical spectra of active galactic nuclei (AGN): broad emission lines with a velocity broadening of a few $10^3~{\rm km~s^{-1}}$ and narrow emission lines with a velocity broadening of a few $10^2~{\rm km~s^{-1}}$. The optical broad and narrow emission lines stem from the so-called broad and narrow emission line regions \citep[BELR and NELR, see][ for a recent review]{net15}, with the former closer \citep[a few light days to weeks,][]{pet04} to the central engine and the latter further away \citep[at least a few parsecs,][]{ben06a,ben06b}. 

In the X-ray band, broad and narrow emission lines are also observed \citep[e.g.,][]{kaa00, ste05, cos07, cos16}, often along with the characteristic narrow radiative recombination continua (RRC) of a photoionized plasma. The spatial extent and overall morphology of the optical and the X-ray NELRs are remarkably similar in a small sample of nearby Seyfert 2 galaxies \citep{bia06}. Nonetheless, there is no conclusive evidence that the optical and X-ray narrow emission features originate from the same photoionized plasma with a multi-phase nature.

In terms of detailed spectral modeling of the narrow emission features in the X-ray band, a two-step approach is commonly used. First, a phenomenological local fit \citep[e.g.,][]{gua07} is performed for individual emission lines and RRC. Typically, the (local) continuum is simply modeled as a (local) power law or spline function plus a model for RRC and an emission line is modeled with a Gaussian or delta profile. The local fit is straightforward and useful, providing primary information like whether a single line is shifted and/or broadened, the temperature of a photoionized plasma (via the width of the RRC), the temperature and density of the plasma (via the line ratios of the He-like triplets), etc. 

With the knowledge obtained from the local fit, the entire spectrum is then modeled with a self-consistent plasma model, or a combination of plasma models \citep[e.g.][]{gua09}. The plasma model can be a photoionized plasma either with a single photoionized component \citep[e.g.,][]{whe15} or multiple photoionized components \citep[e.g.,][]{kin02, arm07, lon08, nuc10, mar11, kal14}, and sometimes a collisional ionized equilibrium (CIE) plasma, which is associated to star formation \citep{gua09} or jets \citep{bia10}. 

For the photoionized plasma modeling, the widely used approach is to simulate a set of photoionized spectra with the Cloudy code \citep{fer17} for a fixed ionizing spectral energy distribution (SED) and for a grid of physical parameters, including the ionization parameter, the line of sight column density and sometimes the plasma number density and microscopic turbulence velocity as well. The number of free parameters and the size of the grid are limited, otherwise, it is computationally too expensive to simulate. Subsequently, the observed spectra are compared with these pre-calculated models, with the best-fit of the physical parameters derived from the closest match. This last step is carried out in a separate spectral fitting package, e.g. XSPEC \citep{arn96}. 

A different approach for the photoionized plasma modeling is used here \citep[see also][]{mao17}. We use SPEX \citep[v3.03.02][]{kaa96} to fit the entire observed spectra on the fly. The advantage of the SPEX code is that it includes an extensive atomic database and self-consistent plasma models, e.g. PION\footnote{The model description and a list of parameters for the PION model can be found in the \href{http://var.sron.nl/SPEX-doc/manualv3.03.00.pdf}{SPEX manual}  (Section 4.29).} for the photoionized plasma, which utilizes the state-of-the-art atomic data \citep{kaa17b}. We refer to \citet{meh16b} for a detailed comparison of the photoionization calculation between SPEX, Cloudy and XSTAR \citep{kal01, bau01}. In each step of the photoionized plasma fitting, the intrinsic SED, thermal equilibrium, ionization balance, level population, transmission, emissivity, and line broadening are calculated in real time to account for the absorption and emission features self-consistently. In short, the photoionization calculation is consistent with the instant ionizing SED and without pre-calculations using grid-defined parameters, thus, more freedom and higher consistency can be achieved in SPEX. 

NGC\,5548 is the archetypal Seyfert 1 galaxy, with broad and narrow emission features across its optical to X-ray spectra \citep[e.g.][]{kor95, chi00, pet02, kaa02, ste05}. The extensive multi-wavelength campaign of NGC 5548 in 2013--2014 \citep{kaa14, meh15} unveiled a special state of the source, where the soft X-ray flux is highly obscured. Such a special state offers a unique opportunity to study the narrow emission lines and radiative recombination continua that were previously hidden by the unobscured continuum \citep{kaa02, ste05, det09}. With a detailed study of the narrow emission features using both the local fit and Cloudy based photoionization modeling, \citet{whe15} interpret the X-ray NELR as a single-phase photoionized plasma that is absorbed by the warm absorber components B+E or A+B+C \citep[see Table S2 of][for the nomenclature]{kaa14}. That is to say, these warm absorber components intervene along the line of sight from the X-ray narrow line emitter to the observer. 

This scenario does not fully agree with the distance estimates of the X-ray NELR and warm absorber components in NGC\,5548. Using the variability of the forbidden line of \ion{O}{vii}, \citet{det09} derived a distance of 1--15~pc for the X-ray NELR, consistent with the optical NELR distance estimate of 1--3~pc \citep{pet13} and joint optical and X-ray NELR distance estimate of $\sim2.4$~pc \citep{lan15}. According to the Cloudy based photoionization modeling, \citet{whe15} found that the X-ray narrow emission lines originate mainly from the illuminated face of the X-ray NELR with a distance of $14$~pc. \citet{ebr16} estimate the distance of the warm absorber components based on variability, with component A and B at least 10~pc away from the central engine and components C to F within 5~pc from the central engine. Nonetheless, based on a spectral analysis using the density sensitive metastable absorption lines, \citet{mao17} found the warm absorber component B is even closer ($<0.23$~pc, $3\sigma$ upper limit) than the optical NELR, rather than further away. \citet{ebr16} constrain the lower limit of the distance of the warm absorber component B based on the non-detection of variability on a timescale of 500 days. However, the authors also pointed out that there are marginal hints of variability at 4 and 60 days. If the variability at a shorter timescale is true, the inferred distance of component B would be much smaller.

Therefore, we reanalyze the high-resolution spectra of NGC\,5548 obtained with the Reflection Grating Spectrometer \citep[RGS,][]{dhe01} aboard \textit{XMM}-Newton \citep{jan01} to provide an alternative interpretation of the X-ray narrow emission features. To be more specific, we attempt to model the soft X-ray emission features with a multi-phase photoionized plasma with no additional absorption by the warm absorber components. \citet{whe15} did not consider this scenario in their study.  

In Section~\ref{sct:obs}, we present the observed RGS spectrum. We describe the detailed spectral analysis in Section~\ref{sct:spec}, including the phenomenological local fit (Section~\ref{sct:local_fit}) and the physical global fit (Section~\ref{sct:global_fit}). The physical global fit is based on the self-consistent photoionization model PION in the SPEX code, with both the single-phase and multi-phase scenarios studied. We discuss the relation between the X-ray and optical narrow emission line region in Section~\ref{sct:rel2opt} and the relation between the X-ray emitter and absorber in Section~\ref{sct:rel2xabs}, respectively. We justify our usage of the unobscured ionizing SED for the X-ray emitter in Section~\ref{sct:ion_sed}. We also point out the abnormally high Ly$\gamma$/Ly$\alpha$ ratio of \ion{N}{vii} and discuss it in terms of a possible charge exchange component in Section~\ref{sct:cx}. The summary can be found in Section~\ref{sct:sum}.

\section{Observations and data reduction}
\label{sct:obs}
The RGS data used here are obtained in two epochs, June 2013 -- February 2014 (PI: J. Kaastra) and January 2016 (PI: G. Kriss), respectively. Data for the first epoch (2013--2014) are taken as part of a large multi-wavelength campaign of NGC\,5548 \citep{kaa14}. There are in total fourteen XMM-{\it Newton} observations ($\sim$50~ks exposure each), where twelve of them are taken between 22 June and 31 July 2013 and the last two are taken in December 2013 and February 2014, respectively. The observation log of 2013--2014 can be found in \citet[][their Table~1]{meh15} and Table~\ref{tbl:obs_log} lists the observation log for 2016.

\begin{table}
\caption{The observation log of NGC\,5548 in 2016 (PI: G.~Kriss) with \textit{XMM}-Newton.}
\label{tbl:obs_log}
\centering
\begin{tabular}{cccccccccccccc}
\hline\hline
\noalign{\smallskip}
Start date & Exp. (ks) & ObsID \\
\noalign{\smallskip} 
\hline
\noalign{\smallskip} 
2016-01-14 & 37 & 0771000101 \\
\noalign{\smallskip} 
2016-01-16 & 34 & 0771000201 \\
\noalign{\smallskip} 
\hline
\end{tabular}
\tablefoot{The observation log in 2013--2014 can be found in Table~1 of \citet{meh15}.}
\end{table}

Details of the RGS data reduction method are similar to those described in \citet{kaa11}. The first-order RGS1 and RGS2 spectra for all observations are stacked for each epoch, with a total exposure of $\sim$770~ks (for 2013--2014) and $\sim$70~ks (for 2016), respectively. The stacked RGS spectrum of 2013--2014 is the same one as that used by \citet{whe15} for the study of the narrow emission features. 

\begin{figure*}
\centering
\includegraphics[width=\hsize, trim={0.3cm 1cm 0.3cm 0.1cm}, clip]{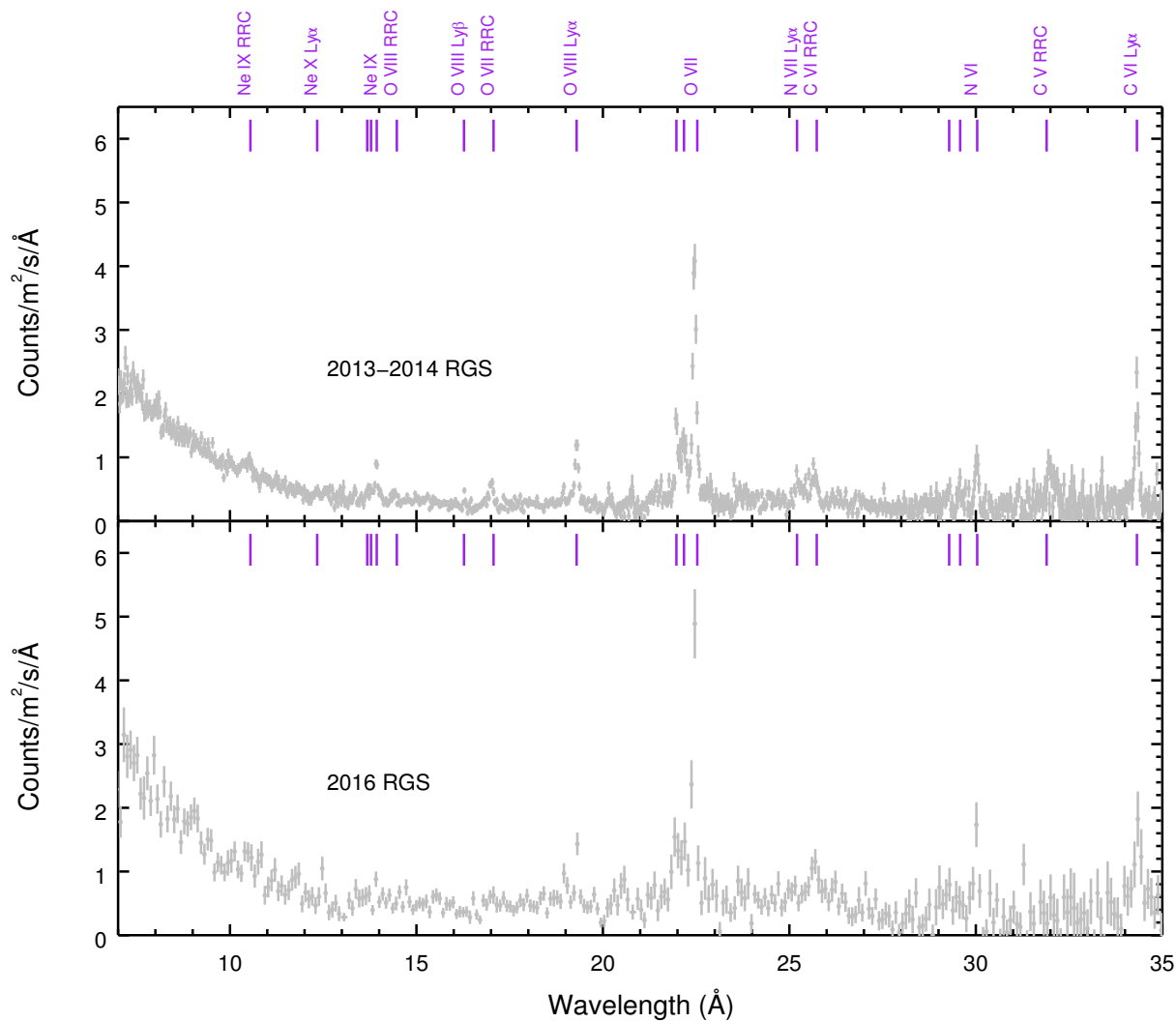}
\caption{The stacked RGS spectra (in the observed frame) in 2013--2014 (the upper panel) and 2016 (the lower panel, rebinned for clarity).}
\label{fig:data_cf_plot}
\end{figure*}

\section{Spectral analysis and results}
\label{sct:spec}
SPEX v3.03.02 is used for the spectral analysis. We use the $C$-statistic throughout this work \citep{kaa17a}. Statistical errors are quoted at 68\% confidence level ($\Delta C$ = 1.0), unless indicated otherwise. The spectral fit is performed in the 7--35 $\angstrom$ wavelength range. Both stacked 2013--2014 and 2016 spectra are optimally binned \citep{kaa16} for the following spectral analysis (Sect.~\ref{sct:spec}). The redshift of the AGN is fixed to $z=0.017175$ \citep{dva91}, as given in the NASA/IPAC Extragalactic Database (NED). 

\subsection{The phenomenological local fit}
\label{sct:local_fit}
In order to check the variability of the narrow emission features between 2013--2014 and 2016, we perform a simple local fit to spectra from the two epochs with a spline continuum model \citep[SPLN in SPEX, see also][]{det08}, Gaussian line models (GAUS) and radiative recombination continua (RRC).

Even in the 770~ks 2013-2014 spectrum, the X-ray broad emission lines are difficult to fit due to their low significance \citep{whe15}. Therefore, in this exercise, we do not include them in both spectra. In the 2016 spectrum, the Ly$\alpha$ line of \ion{O}{viii} and the forbidden line of \ion{O}{vii} can be relatively well constrained, with statistical uncertainties $\lesssim20\%$. The rest of the lines and RRC tabulated in Table~\ref{tbl:cf_local_fit} have statistical uncertainties between 30\% to 100\% in the 2016 spectrum. In Table~\ref{tbl:cf_local_fit}, we compare the intrinsic (unabsorbed) luminosity of these lines and RRC between the two epochs. 

\begin{table}
\caption{The intrinsic (unabsorbed) luminosity of the prominent narrow emission features for the 2013--2014 and 2016 RGS spectra of NGC\,5548.}
\label{tbl:cf_local_fit}
\centering
\begin{tabular}{cccccccccccccc}
\hline\hline
\noalign{\smallskip}
Ion & Line/RRC & $\lambda_0$ & $L_{2013-2014}$ & $L_{2016}$ \\
 & & \AA & $10^{32}$~W & $10^{32}$~W \\
\noalign{\smallskip} 
\hline
\noalign{\smallskip} 
\ion{Ne}{ix} & $1s^2$ & 10.37 & $10.3\pm5.6$ & $23\pm22$ \\
\noalign{\smallskip} 
\ion{Ne}{ix} & He$\alpha$ (f) & 13.70 & $6.7\pm1.2$ & $3.9\pm2.2$ \\
\noalign{\smallskip} 
\ion{O}{vii} & $1s^2$ & 16.79 & $8.7\pm1.6$ & $6.4\pm5.6$ \\
\noalign{\smallskip} 
\ion{O}{viii} & Ly$\alpha$ & 18.97 & $11.2\pm0.7$ & $11.2\pm1.9$ \\
\noalign{\smallskip} 
\ion{O}{vii} & He$\alpha$ (f) & 22.10 & $35.7\pm1.8$ & $39.5\pm5.4$ \\
\noalign{\smallskip} 
\ion{C}{vi} & $1s$ & 25.30 & $13.9\pm6.6$ & $8.9\pm6.7$ \\
\noalign{\smallskip} 
\ion{N}{vi} & He$\alpha$ (f) & 29.53 & $6.4\pm0.9$ & $13.5\pm4.3$ \\
\noalign{\smallskip} 
\ion{C}{vi} & Ly$\alpha$ & 33.74 & $11.7\pm1.2$ & $8.3\pm5.6$ \\
\noalign{\smallskip} 
\hline
\end{tabular}
\tablefoot{$\lambda_0$ is the rest-frame wavelength. }
\end{table}

The most prominent narrow emission lines (the \ion{O}{viii} Ly$\alpha$ and \ion{O}{vii} forbidden lines) have remained constant (at a $1\sigma$ confidence level) in the two epochs June 2013 -- February 2014 and January 2016. This is not totally unexpected, as shown in Table~2 of \citet{det09}, the \ion{O}{vii} line flux is also consistent within 1$\sigma$ for two epochs on a similar timescale, December 1999 -- February 2000 and January 2002. 

On the other hand, some weak features might have varied by a factor of two (the \ion{Ne}{ix} RRC and \ion{N}{vi} forbidden line), but still within a $2\sigma$ confidence level. 

\subsection{The physical global fit}
\label{sct:global_fit}
For the physical global fit (Section~\ref{sct:global_fit}), we first model the high-quality 2013--2014 spectrum with different models (Section~\ref{sct:spec_1314}). Then we simply apply the best-fit model to the low-quality 2016 spectrum (Section~\ref{sct:spec_16}). 

Our global fit includes the following components: (1) The intrinsic broadband spectral energy distribution (SED) of the AGN; (2) The Galactic absorption; (3) The continuum absorption caused by the obscurer; (4) The absorption features caused by the warm absorber; (5) The narrow and/or broad emission features caused by the X-ray emitter. 

The intrinsic SED consists of a Comptonized disk component (COMT in SPEX, for optical to soft X-ray), a power-law component (POW for X-ray), and a reflection component (REFL for the Fe K line and hard X-ray). For the 2013--2014 spectral analysis, all the relevant parameters are fixed to those values given in \citet{meh15}, where the best constraints on these parameters have been obtained from multi-wavelength data. For the 2016 spectral analysis, we fixed the parameters in the COMT component to values that corresponds to the average UVW2 flux in 2016 \citep[the correlations can be found in][]{meh16a}. The power-law and reflection components are allowed to vary in order to match the EPIC-pn data.

The Galactic absorption with $N_{\rm H} = 1.45\times10^{24}~{\rm m^{-2}}$ \citep{wak11} is modeled with the collisional ionisation equilibrium absorption model (HOT) \citep{dpl04, ste05} in SPEX. The electron temperature of the HOT component is fixed to 0.5~eV to mimic the transmission of a neutral gas. 

The continuum absorption caused by the obscurer is modeled with two XABS components \citep{kaa14,meh17}. The line of sight (LOS) hydrogen column densities ($N_{\rm H}$) and absorption covering factors ($f_{\rm cov}$) are allowed to vary in both 2013--2014 and 2016 spectra. The ionization parameters of the two XABS components are treated differently, with the warmer one allowed to vary ($\log\xi\sim-1$) and the cooler one fixed to $-4$ \citep{dge15}. 

The absorption features caused by the warm absorber are accounted for using six PION components. In both the 2013--2014 and 2016 spectra, the hydrogen column densities ($N_{\rm H}$), outflow velocities ($v_{\rm out}$), and microscopic turbulence velocities ($v_{\rm mic}$) of the PION components are fixed to values given in \citet{mao17}, which are obtained by fitting the high quality 2002 \textit{Chandra} grating spectra of NGC\,5548. Nonetheless, the ionization parameters ($\log \xi$) can differ for these three epochs, due to the variability of the obscurer, as well as the changes of the intrinsic SED \citep{cap16}. The ionization parameters are assumed to be proportional to the $1-10^3$~Ryd ionizing luminosity. That is to say, the number density times distance squared ($n_{\rm H} r^2$) in the later two epochs is assumed to be the same as in 2002. 

The narrow and broad emission features caused by the X-ray emitter (sometimes called the warm mirror) are also modeled with PION. Six free parameters of each emission PION component are allowed to vary, including the hydrogen column densities ($N_{\rm H}$), ionization parameters ($\log \xi$), outflow velocities ($v_{\rm out}$), and turbulence velocities ($v_{\rm mic}$) and the emission covering factors ($C_{\rm cov}$, see next paragraph). Additionally, each emission PION component is convolved with a Gaussian velocity broadening model (VGAU in SPEX), with the velocity parameter ($v_{\rm mac}$) free to vary, to account for macroscopic motion. 

Note that there are two covering factors in the PION model. The absorption covering factor $f_{\rm cov}$ is used to model any partial covering in the line of sight (similar to that of the XABS component). The emission covering factor, $C_{\rm cov}=\Omega/4\pi$, corresponds to the normalized solid angle sustained by the emitting region as seen from the central engine. In our modeling, each absorption PION component has fixed $f_{\rm cov} = 1$ and $C_{\rm cov} = 0$, while each emission PION component has fixed $f_{\rm cov} = 0$ and free $C_{\rm cov}\in(0,~1)$. Furthermore, the ionizing SED for the absorption PION components is the obscured SED, as the obscurer locates between the warm absorber and the central engine \citep{kaa14}. Nevertheless, the ionizing SED received by the emission PION components is assumed to be unobscured, and we will discuss this in detail in Section~\ref{sct:ion_sed}. Unless indicated otherwise, the proto-solar abundances of \citet{lod09} are used for all plasma models (HOT, XABS and PION).  

\subsubsection{The 2013--2014 RGS spectrum}
\label{sct:spec_1314}
\begin{figure}
\centering
\includegraphics[width=\hsize, trim={0.5cm, 1cm, 1cm, 0.1cm}, clip]{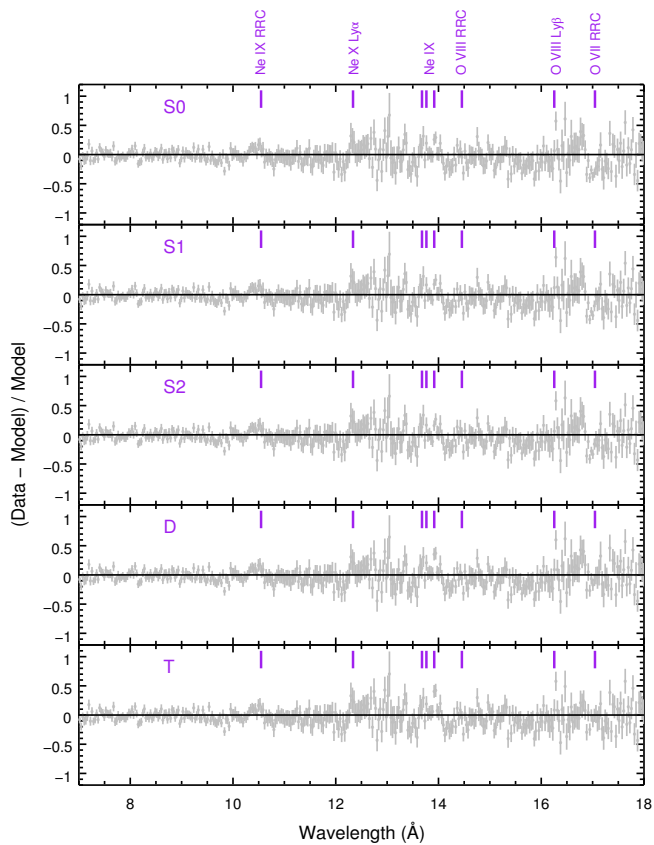}
\caption{The fit residuals of the physical models in the 7--18~\AA\ band.}
\label{fig:dchi_cf_plot_part1}
\end{figure}

\begin{figure}
\centering
\includegraphics[width=\hsize, trim={0.5cm, 1cm, 1cm, 0.1cm}, clip]{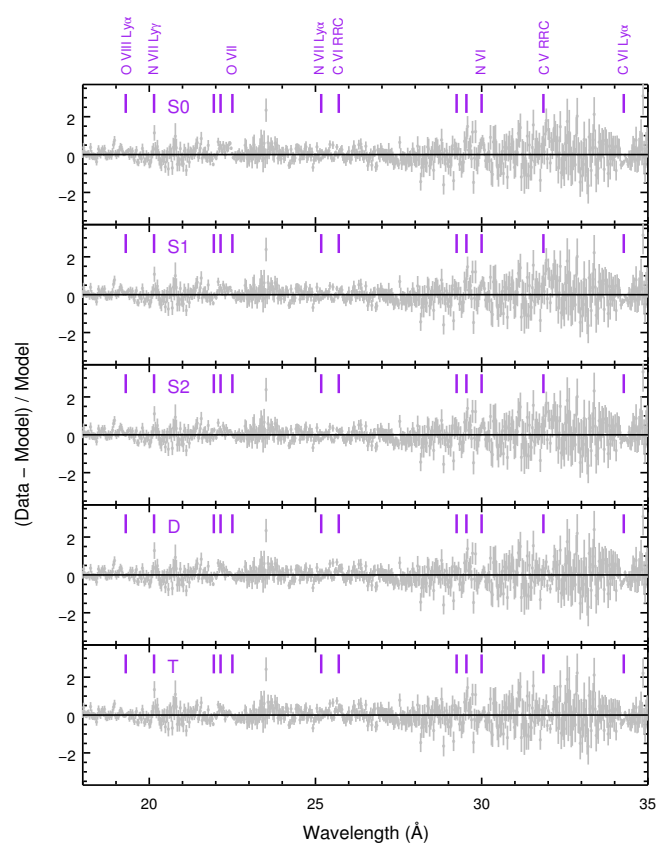}
\caption{As Figure~\ref{fig:dchi_cf_plot_part1} in the 18--35~\AA\ band.}
\label{fig:dchi_cf_plot_part2}
\end{figure}

Five different models are used for the emission features in the RGS band. We start with a single-phase photoionized emitter (denoted as Model S0), which can reproduce well some of the observed narrow emission features, but not all. In particular, Model S0 fails to match the RRC of \ion{O}{vii} ($\sim$17~\AA) and \ion{C}{vi} ($\sim$32~\AA), the forbidden lines of \ion{N}{vi} ($\sim30$~\AA), etc. The fit residuals can be found the in top panels in Figures~\ref{fig:dchi_cf_plot_part1} and \ref{fig:dchi_cf_plot_part2}. The residuals indicate that either additional absorption is required  or the X-ray emitter has at least two emission components. 

Following \citet{whe15}, we apply absorption caused by the warm absorber components B+E (Model S1) or A+B+C (Model S2) to the single-phase photoionized emitter. As expected, for both models, the $C$-stat is significantly improved (Table~\ref{tbl:cf_global_fit}). Model S2 yields more consistent parameters with \citet{whe15}, i.e. a mildly ionized ($\log \xi\sim1-1.5$), blueshifted ($v_{\rm out}\sim-300~{\rm km~s^{-1}}$), turbulent ($v_{\rm mic}\sim 200~{\rm km~s^{-1}}$) plasma with $N_{\rm H}\sim10^{25-26}~{\rm m^{-2}}$. The different values of the best-fit parameters and $C$-stat between the present work and \citet{whe15} are not unexpected. We use here a narrower wavelength range (7--35~\AA), ignoring the 5.7--7~\AA\ and 35--38.2~\AA\ wavelength ranges, where the noise is relatively large and no strong lines or RRC are expected. Additionally, due to the different atomic database and calculation used by Cloudy and SPEX, fitting the same high-resolution spectra can lead to best-fit parameters that differ by 10--40\% \citep{meh16b}. 

\begin{table*}
\caption{Photoionized models of the emission features for NGC\,5548.}
\label{tbl:cf_global_fit}
\centering
\begin{tabular}{cccccccccccccc}
\hline\hline
\noalign{\smallskip}
Model & Comp. & $N_{\rm H}$ & $\log~(\xi)$ & $v_{\rm mic}$ & $v_{\rm out}$ & $C_{\rm cov}$ & $v_{\rm mac}$ & $C$-stat & d.o.f. \\
 & & (${\rm 10^{25}~m^{-2}}$) & (${\rm 10^{-9}~W~m}$) & (${\rm km~s^{-1}}$) & (${\rm km~s^{-1}}$) & (\%) & (${\rm km~s^{-1}}$) & &  \\
\noalign{\smallskip} 
\hline
\noalign{\smallskip} 
S0 & EM~1 & $12.7\pm1.9$ & $1.18\pm0.02$ & $460\pm30$ & $-100\pm20$ & $2.4\pm0.3$  & $<210$ & 1774 & 920 \\ 
\noalign{\smallskip} 
\hline
\noalign{\smallskip} 
S1 & EM~1 & $12.4\pm1.4$ & $1.23\pm0.02$ & $420\pm30$ & $-190\pm10$ & $2.9\pm0.3$  & $<250$ & 1652 & 920 \\  
\noalign{\smallskip} 
\hline
\noalign{\smallskip} 
S2 & EM~1 & $6.3\pm1.2$ & $1.20\pm0.02$ & $200\pm50$ & $-277\pm4$ & $6.3\pm0.9$  & $400\pm40$ & 1569 & 920 \\ 
\noalign{\smallskip} 
\hline
\noalign{\smallskip} 
\multirow{2}{*}{D} & EM~1 & $14.7\pm0.1$ & $1.30\pm0.02$ & $520\pm40$ & $-49\pm6$ & $1.8\pm0.6$  &  $<160$ &  \multirow{2}{*}{1567} & \multirow{2}{*}{914} \\ 
 & EM~2 & $19.3\pm1.3$ & $0.14\pm0.04$ & $250\pm60$ & $-410\pm50$ & $0.60\pm0.04$ & $<220$ &  \\ 
\noalign{\smallskip} 
\hline
\noalign{\smallskip} 
\multirow{3}{*}{T} & EM~1 & $9.7\pm1.3$ & $1.31\pm0.02$ & $400\pm{30}$ & $-47\pm4$ & $2.2\pm0.2$ & $<100$ &  \multirow{3}{*}{1525} & \multirow{3}{*}{910}\\ 
 & EM~2 &  $30\pm7$ & $0.13\pm0.05$ & $<280$ & $-420\pm30$ & $0.41\pm0.07$ & $260\pm80$ &  \\
 & EM~3 &  $23\pm6$ & $1.24\pm0.07$ & 100 (f) & 0 (f)  & $0.5\pm0.3$ & $7400\pm1100$ &  \\ 
\noalign{\smallskip} 
\hline
\end{tabular}
\tablefoot{The emission covering factor ($C_{\rm cov}=\Omega/4\pi$) refers to normalized the solid angle ($\Omega$) subtended with respect to the central engine. The expected $C$-stat is $935\pm43$ for all the models. The degree of freedom (d.o.f.) is for the RGS band. For EM~3 in Model T, $v_{\rm mic}$ and $v_{\rm out}$ are frozen.}
\end{table*}

\begin{figure*}
\centering
\includegraphics[width=\hsize, trim={0.5cm, 2.0cm, 0.5cm, 0.7cm}, clip]{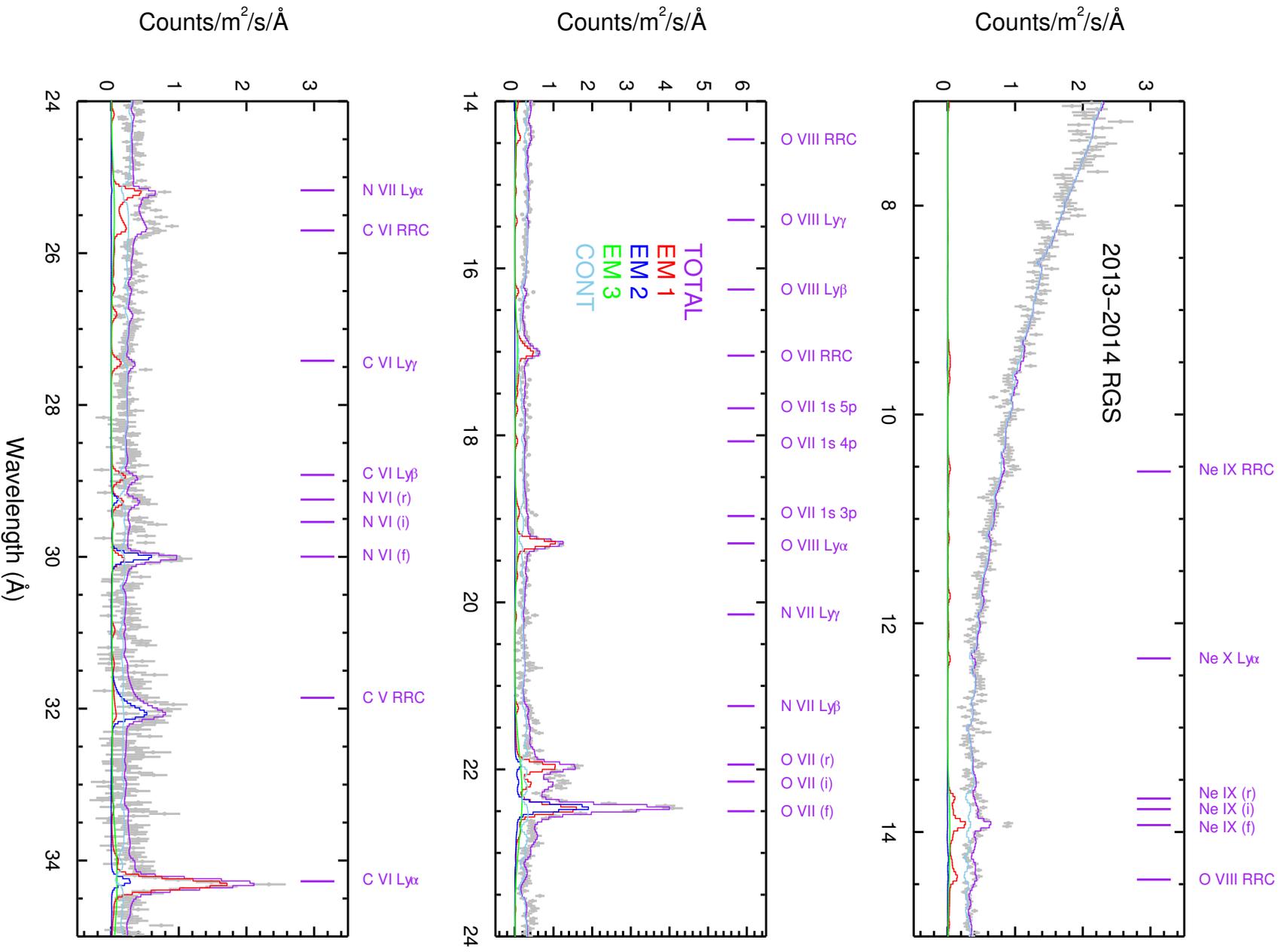}
\caption{The best-fit to the 2013--2014 RGS spectrum with three emission components (Model T) for the narrow and broad emission features.}
\label{fig:spec_zoom_plot}
\end{figure*}

Alternatively, we also try a model with two emission components (Model D) to account for the narrow emission features. The hotter ($\sim4.1$~eV) component (EM~1) has a higher luminosity ($\sim1.2\times10^{34}$~W in the 7--35~\AA\ band) and a larger radiation to gas pressure ratio\footnote{The radiation to gas pressure ratio, also known as the pressure form of the ionization parameter, $\Xi=L/(4\pi r^2 n_{\rm H} c kT)=\xi/(4\pi c kT)$, where $L$ is the 1--1000 Ryd ionizing luminosity, $r$ the distance of the slab, $n_{\rm H}$ the hydrogen number density, $k$ the Boltzmann constant and $T$ the electron temperature \citep{kro81}.} ($\Xi\sim8$). The cooler ($\sim1.6$~eV) component (EM~2) has a lower luminosity ($\sim3.4\times10^{33}$~W in the 7--35~\AA\ band) and a smaller radiation to gas pressure ratio ($\Xi\sim1.4$). The $C$-stat is improved significantly when comparing Model D to Model S0 ($\Delta C\sim-200$, to be compared with the root-mean-square deviation 43 of the expected $C$-stat). But it is negligible ($\Delta C=-2$) when compared to Model S2, as Model D fits better the RRC of \ion{O}{vii}, yet slightly worse the Ly$\alpha$ line of \ion{C}{vi}. Adding a third emission component (EM~3 in Model T) with a very broad line profile can further improve the $C$-stat ($\Delta C\sim-40$), but not significant compared with the root-mean-square deviation 43 of the expected $C$-stat. The best-fit (Model T) to the 7--35~\AA\ wavelength range RGS data is shown in Figure~\ref{fig:spec_zoom_plot}.

The narrow emission lines and RRC of \ion{Ne}{x} (H-like) and \ion{Ne}{ix} (He-like) are underestimated with all the models above. However, adding another emission component with a narrow line profile (including both the turbulent and Gaussian velocity broadening) does not improve the $C$-stat. An ad hoc solution is to use a super-solar Ne abundance ($1.7\pm0.2$) for the highly ionized emission component (EM~1), which fits the Ne emission features well and improves the $C$-stat ($\Delta C\sim-30$). 

\begin{figure}
\centering
\includegraphics[width=\hsize, trim={0.0cm 0.5cm 0.5cm 0.5cm}, clip]{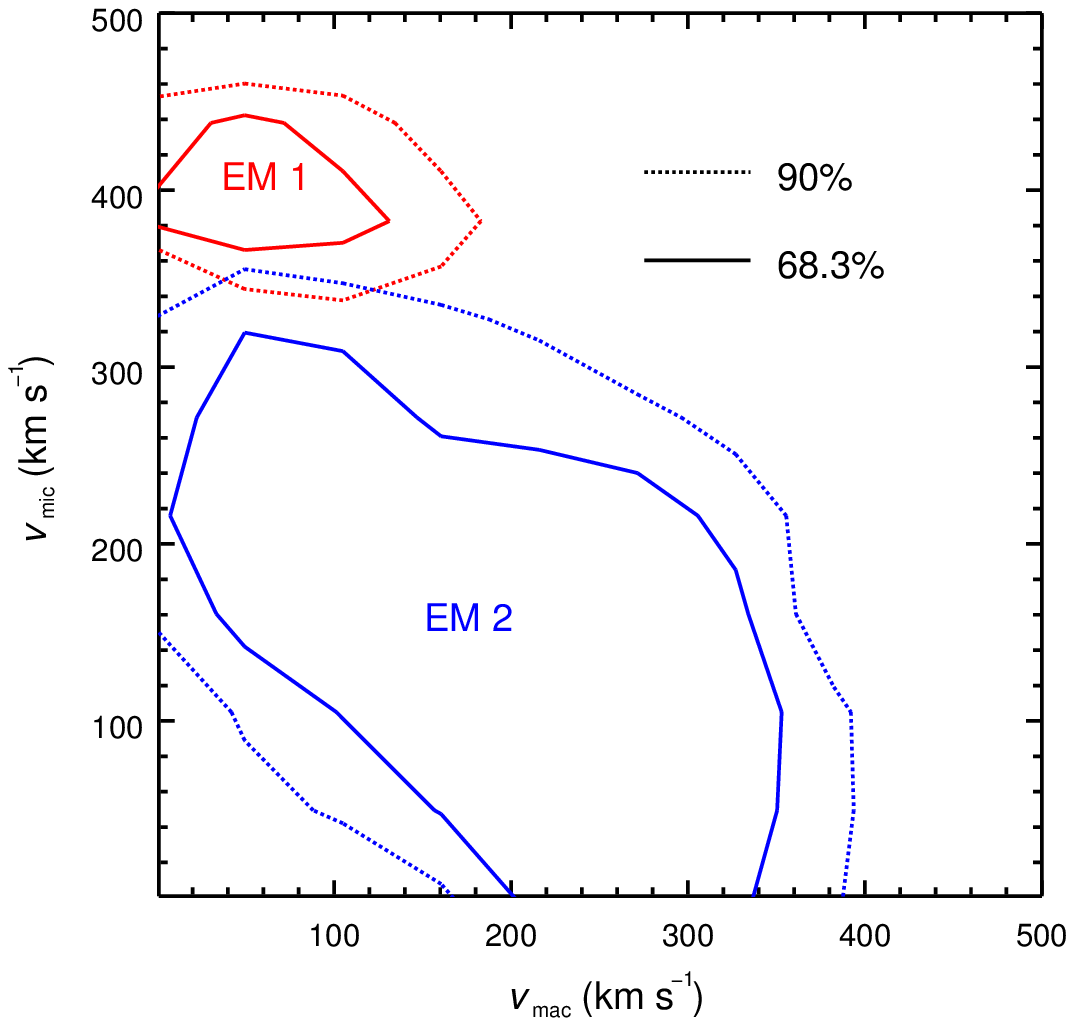}
\caption{The confidence level contours for the microscopic turbulence velocities ($v_{\rm mic}$) and macroscopic motion velocities ($v_{\rm mac}$) of EM~1 (in red) and 2 (in blue) in Model T. The solid and dotted contours refer to $68.3\%$ (or $\Delta C=2.30$) and $90\%$ (or $\Delta C=4.61$), respectively. }
\label{fig:vmic_vmac_contour_plot}
\end{figure}

In all the models above, the microscopic turbulence velocity ($v_{\rm mic}$) and the macroscopic motion ($v_{\rm mac}$) velocity are highly degenerate. That is to say, the $1\sigma$ uncertainties on $v_{\rm mic}$ and $v_{\rm mac}$ are underestimated ($\lesssim30$\%) when considering either one of two parameters alone, as in Table~\ref{tbl:cf_global_fit}. Figure~\ref{fig:vmic_vmac_contour_plot} shows the confidence level contours of EM~1 and 2 in Model T. In this case, $v_{\rm mac}$ is less well constrained than $v_{\rm mic}$. The former is merely constrained by the velocity broadening of the line, while the latter puts an extra limit on the optical depth of the line. When $v_{\rm mac}$ is negligible, $v_{\rm mic}$ accounts for both the optical depth and velocity broadening. On the other hand, if $v_{\rm mic}$ is negligible (not the case for EM~1 in Model T), $v_{\rm mac}$ dominates the line broadening.

\subsubsection{The 2016 RGS spectrum}
\label{sct:spec_16}
\begin{figure*}
\centering
\includegraphics[width=\hsize, trim={0.3cm 1cm 0.3cm 0.1cm}, clip]{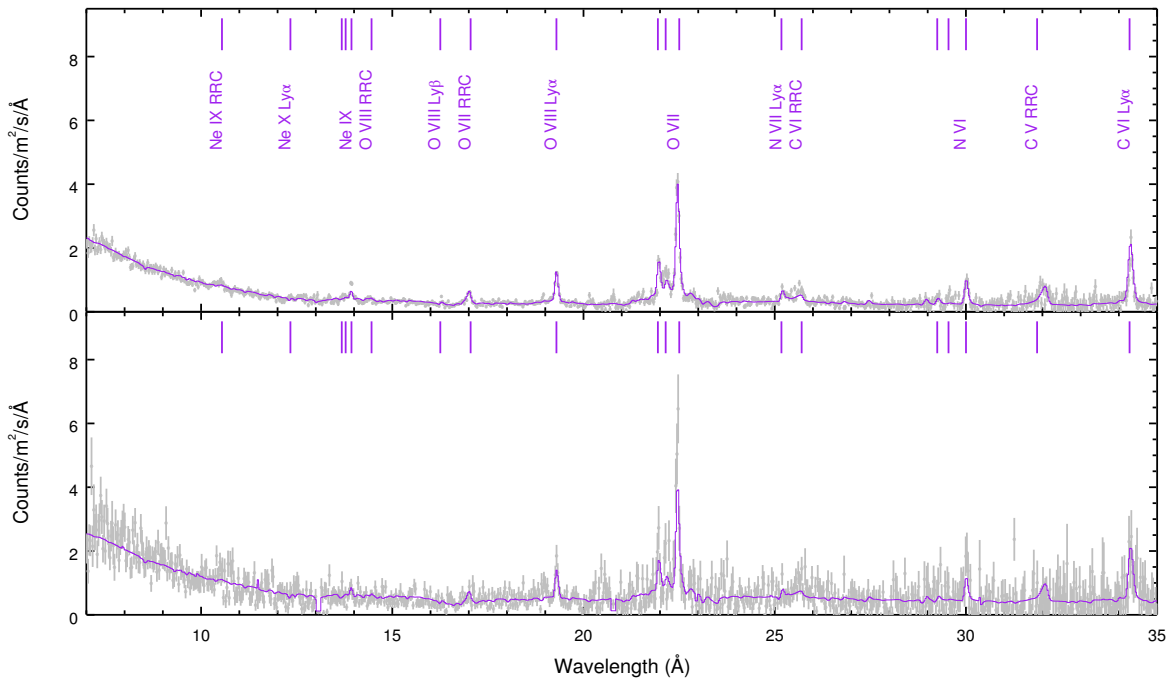}
\caption{With the model of three emission components (Model T), the best-fit to the RGS spectra (in the observed frame) in 2013--2014 (the upper panel) and 2016 (the lower panel, rebinned for clarity).}
\label{fig:spec_cf_plot}
\end{figure*}

In the global fit of the 2016 spectrum, the power-law component, the reflection component, the two XABS components (Section~\ref{sct:global_fit}) are allowed to vary. Assuming the X-ray emitter remains unchanged between the two epochs (Section~\ref{sct:local_fit}), with Model T for the narrow and broad emission features, the best-fit $C$-stat over expected $C$-stat ratio is 1443/1304 for the RGS band. Contrary to our simple assumption that the narrow emission features are constant between the two epochs, we do find that the forbidden lines of \ion{O}{vii} and \ion{N}{vi} are underestimated in our model when applied to the 2016 spectrum. These results are similar to what we found with the simple phenomenological fit (Section~\ref{sct:local_fit}). Figure~\ref{fig:spec_cf_plot} shows the comparison of the best-fit model to both spectra in 2013--2014 and 2016. 

\section{Discussion}
\label{sct:dis}
\subsection{In relation to the optical NELR}
\label{sct:rel2opt}
From the variability of the optical narrow emission line [\ion{O}{iii}] 4959~\AA\ and 5007~\AA, \citet{pet13} deduced that the optical narrow emission line region (NELR) has a radius of 1--3~pc with $n_{\rm e}\sim10^{11}~{\rm m^{-3}}$. A variability study of the X-ray narrow emission line [\ion{O}{vii}] 22.10~\AA\ yields an X-ray NELR distance of 1--15 pc \citep{det09}, which is consistent with the optical NELR. We cannot constrain the density and distance in the X-ray spectra. Although the He-like triplet of \ion{O}{vii} is prominent, the associated density diagnostic is not effective, since the forbidden to intercombination line ratio of \ion{O}{vii} is strongly sensitive to the density only when $n_{\rm e}\sim10^{15-18}~{\rm m^{-3}}$ \citep[][their Fig.~10]{por00}, which is too high for the NELR.

The asymmetric line profiles of the [\ion{O}{iii}] lines indicate that the NELR has an outflowing component, with ${\rm HWHM_{\rm blue}} / {\rm HWHM_{\rm red}} \simeq 2.12$ with an outflow velocity of $\sim460~{\rm km~s^{-1}}$ \citep{pet13}. Similarly, the line profile of \ion{O}{vii} observed in the X-ray band is also asymmetric \citep[see Fig.~3 in][]{whe15}. This can also be seen in our Models D and T, where two emission components with different outflow velocities are required. The emission component (EM 2) that dominates the forbidden lines of \ion{O}{vii} and \ion{N}{vi} (but not that of \ion{Ne}{ix}, see Figure~\ref{fig:spec_zoom_plot}) is blueshifted ($\sim-400~{\rm km~s^{-1}}$). The other emission component (EM 1), which dominates the Ly$\alpha$ lines of \ion{O}{viii}, \ion{N}{vii} and \ion{C}{vi} (Figure~\ref{fig:spec_zoom_plot}), has a negligible outflow velocity. Such divergent kinematic behaviour can also be found in the phenomenological fit by \citet[][their Table 6]{whe15}. 

We point out that \citet{pet13} attribute the velocity broadening to the virial motion of the gas and no microscopic turbulence velocity broadening is taken into account. As pointed out by \citet{kra07}, it is possible that microturbulence is present in the optical NELR of NGC\,5548. Furthermore, if turbulence dissipates within the NELR, the plasma can be heated in excess of the temperature corresponding to photoionization. In our photoionization modeling of the X-ray narrow emission features, nondissipative microturbulence is taken into account. That is to say, the microscopic and macroscopic velocity broadening are both taken into account, but no extra heating is used to solve the thermal equilibrium. 

In reality, it is possible that turbulence dissipates within the NELR, in practice, it is difficult to model the emergent spectrum. On one hand, the extra heating can be included via the external heating option in the PION model\footnote{In \citet{kra07}, the dissipative heating is introduced via the additional heating terms in the Cloudy modeling.}, if we know the scale length over which the turbulence dissipates, the plasma mass density and turbulence velocity of the photoionized plasma \citep[][their Eq.~1]{bot02a}. The dissipation scale length and direction, as well as the mass density, can only be assumed in our analysis. On the other hand, the line broadening might appear to be different for the resonance and forbidden lines. Considering a slab with suitable ionization parameter for the resonance and forbidden lines, assuming the turbulence dissipation direction is along the line of sight from the central engine toward the observer, and the dissipation length scale equals the size of the slab, the observed broadening of the forbidden lines is the integrated result of velocity broadening across the entire slab because they are optically thin. In contrast, due to the large optical depth, the observed resonance lines are only those escaping from the skin ($\tau\sim1$) of the far (with respect to the central engine) side of the slab, therefore, they are less broadened since turbulence converts to heating at the far side. In short, in order to model the emergent spectrum of a dissipative turbulent photoionized plasma, several assumptions are required, and the line broadening effect for resonance and forbidden lines are not trivial. 

The bottom line is that the total velocity broadening of the narrow emission lines is $300-500~{\rm km~s^{-1}}$ for individual emission components, well in excess of the thermal broadening. Both the microscopic turbulence and macroscopic motions can contribute to the total velocity broadening, but the two velocities are highly degenerate (Figure~\ref{fig:vmic_vmac_contour_plot}).

Furthermore, regardless of the number of emission components we used for the X-ray narrow emission features, the best-fit emission covering factor $C_{\rm cov}=\Omega/4\pi$ is rather small, with $\sim6\%$ for Model S2, and $2-3\%$ for the other models. This implies a compact geometry for the X-ray NELR. We caution that the photoionized plasma is assumed to be uniform with the PION model. If the NELR is clumpy, then the true covering factor can be larger. On the other hand, the optical NELR of NGC\,5548 is also found to be compact \citep[$C_{\rm cov}\sim11\%$,][]{kra98}. 

To summarize, in this work, the X-ray narrow emission features show asymmetric line profiles, and the underlying photoionized plasma is turbulent and compact in size. Similar results have been reported in previous optical studies of the NELR \citep{kra98, kra07, pet13}. Moreover, distance estimates (from previous studies) of the X-ray and optical NELR indicate that the two regions might be co-located, i.e. a few parsec away from the central engine \citep{det09,pet13,lan15}. These similarities further suggest that the X-ray and optical NELR might be the same multi-phase photoionized plasma that manifests its emission in different energy bands. This interpretation has actually been established with the radiation pressure confinement (RPC) model \citep[e.g.][]{ste14}. Of course, the RPC model is more sophisticated, where the multi-phase photoionized plasma has a range of number densities and ionization parameters. We only have two uniform slabs for the X-ray narrow emission features. But we do notice that the radiation to gas pressure ratio decreases from the highly ionized component EM~1 to the lowly ionized component EM~2 (Section~\ref{sct:spec_1314}), which agrees with the RPC model.

Interestingly, based on the measurement of the metastable absorption lines (\ion{C}{iii} 1175~\AA\ and \ion{Si}{iii} 1298~\AA) in the UV band, \citet{ara15} determined the number density ($\log n_{\rm e}=10.8\pm0.1~{\rm m^{-3}}$) and distance ($\sim3.5$~pc) of the UV absorption component 1, which are also similar to those of the optical NELR.

\subsection{In relation to the X-ray warm absorber}
\label{sct:rel2xabs}
As can be seen from Table~\ref{tbl:cf_global_fit}, the single-phase emission component fully covered by the warm absorber components A+B+C \citep[][i.e. Model S2 in this work]{whe15} and the two-phase emission components with no overlying warm absorber components (Model D) yield comparable fit statistics. Nevertheless, the underlying geometries of the two scenarios are rather different.
 
\begin{figure}
\centering
\includegraphics[width=\hsize, trim={0.3cm 0.1cm 0.3cm 0.1cm}, clip]{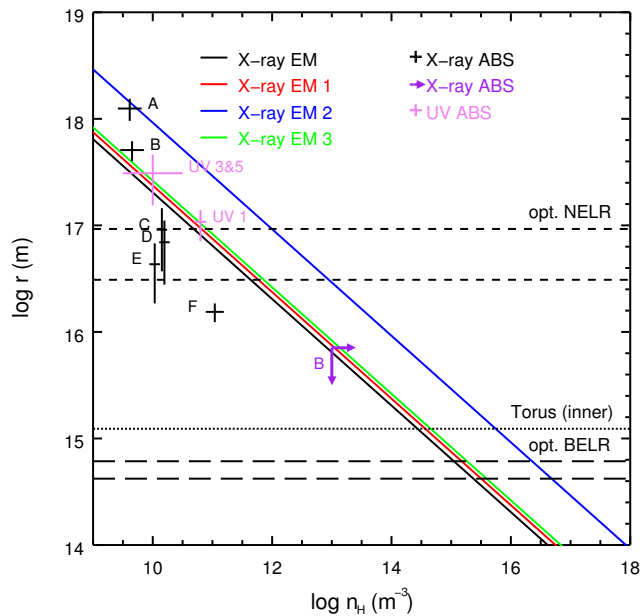}
\caption{Distance and number density of optical, UV and X-ray emission and/or absorption components in NGC\,5548. The solid diagonal lines are the distance and density relations (i.e. $\log \xi = L / n_{\rm H}r^2$) of the X-ray emission components \citep[EM for][and EM~1--3 for Model T]{whe15}. The black pluses ($1\sigma$ uncertainties) refer to the six warm absorber components \citep[via a timing analysis,][]{ebr16}. The purple arrows are the $3\sigma$ upper (distance) and lower (density) limits of the warm absorber component B \citep[via a spectral analysis,][]{mao17}. The pink pluses are for the UV absorption components \citep[][]{ara15}. The horizontal short dashed lines refer to the optical narrow emission line region \citep{pet13}. The horizontal long dashed lines refer to the optical broad emission line region \citep{kas00, bot02b}. The horizontal dotted line indicates the inner edge of the torus \citep{sug06}.}
\label{fig:nddist_plot}
\end{figure}

In the first scenario (Model S2), the warm absorber components (A to C) are located between the X-ray emitter and the observer, with the warm absorber components fully covering the emission component. Given that the X-ray NELR is compact in size (Section~\ref{sct:rel2opt}), full covering by the warm absorber components is possible, as long as the warm absorber components are located further away than the emission component. \citet{ebr16} estimate the distances of the warm absorber components via a variability study. The warm absorber components A and B are at least 10~pc away from the central engine, while component C is between 1 and 5~pc (Figure~\ref{fig:nddist_plot}).

Nevertheless, a distance estimation based on the density sensitive metastable absorption lines found a $3\sigma$ upper limit of 0.23~pc for the warm absorber component B \citep{mao17}. If this is the case, and assuming the X-ray NELR has a distance of 1--3~pc, the warm absorber component B cannot absorb the NELR. The second scenario (Model D), with no overlying absorption, does not suffer from the logical difficulty, if the warm absorber component B is closer than the X-ray NELR. 

Distance (and also density) estimation using either timing or spectral analysis is challenging. Timing analyses usually suffer from low cadence and sometimes the signal-to-noise ratio is not enough to claim a significant change. On the other hand, spectral analyses (in the X-ray band) suffer from low spectral resolution and limited photon collecting area in the relevant energy band with current grating spectrometers. Future studies with either an intensive monitoring program or a spectral analysis with the next generation spectrometers \citep[e.g. \textit{Arcus,}][]{smi16} are required to better constrain the number density and distance of the warm absorber \citep{kaa17b}.

We also compare the parameters of the emission and absorption components. We list the parameters for selected emission and absorption comparison in Table~\ref{tbl:cf_em_abs}. The term $n_{\rm H}r^2$ of the X-ray emission component 1, (X-ray) warm absorber component B and UV absorption component 1 are comparable ($\sim10^{44}~{\rm m^{-1}}$). The distances (thus number densities) of EM~1 is consistent with UV~1, but not with X-ray component B (Figure~\ref{fig:nddist_plot}). The other parameters ($N_{\rm H}$, $v_{\rm out}$, and $v_{\rm mic}$) of these emission and absorption components do not agree with each other. For the X-ray absorption component A and X-ray emission component 2, three parameters ($n_{\rm H}r^2$, o$v_{\rm out}$ and $v_{\rm mic}$) are roughly of the same order of magnitude, but the hydrogen column density and distance (thus number density) are not consistent with each other. 

The warm absorber components C to E have similar distances as the X-ray NELR (Figure~\ref{fig:nddist_plot}), but the inferred number densities are lower by an order of magnitude, while component F has smaller distance and number density than the X-ray NELR \citep{ebr16}. The UV absorption components 3 and 5 are slightly further away (5--15~pc) than the X-ray NELR, while the distances of components 2, 4, and 6 are not well constrained \citep{ara15}. 

In short, the emission components are not the counterpart of the UV/X-ray absorption components outside the line of sight. 

\begin{table*}
\caption{Comparing the parameters of selected emission and absorption components.}
\label{tbl:cf_em_abs}
\centering
\begin{tabular}{cccccccccccccc}
\hline\hline
\noalign{\smallskip}
Comp. & $N_{\rm H}$ & $n_{\rm H} r^2$ & $v_{\rm mic}$ & $v_{\rm out}$ & E.M. & $r$ \\
 & ($10^{25}~{\rm m^{-2}}$) & ($10^{44}~{\rm m^{-1}}$) & (${\rm km~s^{-1}}$) & (${\rm km~s^{-1}}$) & (${\rm 10^{70}~m^{-3}}$) & (pc)  \\
\noalign{\smallskip} 
\hline
\noalign{\smallskip} 
EM~1 & $9.7\pm1.3$ & $5.7\pm0.3$ & $400\pm30$ & $-47\pm4$ & $1.8\pm0.6$ & 1--3$\ddagger$  \\
\noalign{\smallskip} 
B & $0.69\pm0.09$ & $5.2\pm0.8$ & $49\pm14$ & $-550\pm40$ & -- -- & $<0.23$ or 13--20$\dagger$  \\
\noalign{\smallskip} 
UV~1 & $3.2_{-1.2}^{+4.7}$ & $7.4_{-4.5}^{+4.3}$ & -- -- & $-1160$ & -- -- & $3.5_{-1.2}^{+1.0}$  \\
\noalign{\smallskip} 
\hline
\noalign{\smallskip} 
EM~2 & $30\pm7$ & $87\pm11$ & $<280$ & $-420\pm30$ & $16\pm10$ & 1--3$\ddagger$ \\
\noalign{\smallskip} 
A & $0.26\pm0.08$ & $36\pm12$ & $150\pm30$ & $-570\pm40$ & -- -- & 31--50$\dagger$ \\
\hline
\end{tabular}
\tablefoot{Results from this work and \citet[][$\ddagger$]{pet13} are used for the narrow emission components (EM~1 and 2 in Model T). The warm absorber components (A and B) are from \citet{mao17} and \citet[][$\dagger$]{ebr16}. The UV absorption component 1 is from \citet{ara15}. We do not compare the ionization parameter, because it can vary due to the change in the ionizing luminosity. The emission measure (E.M.) is calculated via $ n_{\rm e}~n_{\rm H}~4\pi C_{\rm cov}r^2~N_{\rm H} / n_{\rm H}$.}
\end{table*}

\subsection{Unobscured SED for the X-ray emitter}
\label{sct:ion_sed}
Although we have witnessed significant spectral changes since June 2011\footnote{There were no observations of NGC 5548 between August 2007 and June 2012 with \textit{XMM}--Newton, \textit{Chandra}, \textit{Suzaku}, or \textit{Swift}.} \citep{kaa14}, we argue that using the unobscured SED for the X-ray emitter \citep[same in][]{whe15} is reasonable given the following effects.

First, there is a geometric effect. A possible geometry is shown in Fig.4 of \citet{kaa14}, where the obscurer appears along our line of sight, yet only a small part is in-between the narrow line region and the black hole and the accretion disk. 

Second, there is a time delay effect. It is also possible that the X-ray emitter has not yet responded to the changes in the ionizing SED, whether it is obscured or not. The delay timescale ($\tau_{\rm delay}$) is the sum of the light travel time ($\tau_{\rm lt}$) and the recombination time ($\tau_{\rm rec}$). The recombination timescale mainly depends on the number density ($n_{\rm H}$) of the plasma \citep{kro95, nic99}. The higher the plasma density, the shorter the recombination time. In our photoionization modeling, the recombination timescale for \ion{O}{vii} is roughly on the order of $10^{16}~(n_{\rm H}/{\rm m^{-3}})^{-1}~{\rm s}$. The light travel time is simply $3.26~(r_{\rm X}/{\rm pc})~{\rm yr}$, i.e., $10^8~(r_{\rm X}/{\rm pc})~{\rm s}$, where $r_{\rm X}$ is the distance of the X-ray NELR. As long as the number density of the photoionized plasma is $\gtrsim10^{10}~{\rm m^{-3}}$ and the distance of the X-ray NELR $\gtrsim1$~pc, the delay time ($\gtrsim$3.3 yr) is longer than the time separation ($\sim$2.6~yr) between our two epochs (2013--2014 and 2016).

Third, there is a low-density effect. When the density of the photoionized plasma is sufficiently low, the recombination timescale ($\tau_{\rm rec}$) is orders of magnitude higher than the variability timescale ($\tau_{\rm var}$), and the plasma is in a quasi-steady state. In other words, the ionization balance of the plasma varies slightly around the mean value corresponding to the mean ionizing flux level over time \citep{nic99, kaa12, sil16}. 

\subsection{A charge exchange component?}
\label{sct:cx}
In our physical global fit (Section~\ref{sct:global_fit}) to the stacked RGS spectrum in 2013--2014, we find that the \ion{N}{vii} Ly$\gamma$\footnote{19.83~\AA\ in the rest-frame and 20.16~\AA\ in the observed frame.} narrow emission line is abnormally high for a photoionized plasma (all panels in Figure~\ref{fig:dchi_cf_plot_part2}). When we include a Gaussian profile for the \ion{N}{vii} Ly$\gamma$ narrow emission line, the $C$-stat is significantly improved with $\Delta C=-23$, at the price of two degrees of freedom. The best-fit line luminosity is $(1.9\pm0.5)\times10^{32}~{\rm W}$, and the \ion{N}{vii} Ly$\gamma$/Ly$\alpha$ ratio is $0.7_{-0.3}^{+1.3}$. Such a high Ly$\gamma$/Ly$\alpha$ ratio cannot be explained by a photoionized plasma, nor a collisionally ionized plasma, but it can be obtained with a charge exchange plasma \citep[e.g.][]{gu15}. 

It is possible for charge exchange events to occur if a warm ($T\sim 10^{5-6}$~K) outflow runs into the cold ($T\lesssim 1800$~K) torus region. Assuming a constant radial velocity of $\sim 300~{\rm km~s^{-1}}$, if the outflow is short-lived ($\lesssim10^{2-3}$~yr), only gas arising beyond \citep{blu05} the broad line region can reach the torus region. If the outflow is long-lived, then outflows arising from the accretion disk \citep{pro00} can also reach the torus region. In fact, putative charge exchange emission features at 1223.6~\AA, 1242.4~\AA, and 1244.0~\AA\ (for \ion{Ne}{x} and \ion{S}{xv}) in the UV spectrum of \object{NGC\,1275} have been reported in \citet{gu17}. 

In our case, using a charge exchange component \citep[the CX model in SPEX,][]{gu16}, we can reproduce a \ion{N}{vii} Ly$\gamma$/Ly$\alpha$ ratio of $\sim$0.7, but the same CX plasma would also produce high Ly$\gamma$/Ly$\alpha$ for \ion{C}{vi} and Ly$\delta$/Ly$\alpha$ for \ion{O}{viii} ratios ($\sim$0.5), which are not found in the observed spectra. Therefore, we are not convinced that the \ion{N}{vii} Ly$\gamma$/Ly$\alpha$ ratio alone can validate the presence of a charge exchange plasma in NGC\,5548.

\section{Summary}
\label{sct:sum}
We reanalyze the high-resolution spectrum of the archetypal Seyfert 1 galaxy NGC\,5548 obtained with the Reflection Grating Spectrometer aboard \textit{XMM}-Newton in 2013--2014 (770~ks), and analyze the spectrum in 2016 (70~ks). The main results are summarized as follows. 

\begin{enumerate}
    \item The most prominent emission lines (the \ion{O}{vii} forbidden line and \ion{O}{viii} Ly$\alpha$ line) are consistent (at a 1 $\sigma$ confidence level) between 2013--2014 and 2016. This is not totally unexpected, as the non-variability at such a timescale has been reported previously.  
    \item The X-ray narrow emission line region (NELR) can be modeled as a two-phase photoionized plasma without further absorption by the warm absorber. This is an alternative to the previous interpretation of the NELR as a single-phase photoionized plasma absorbed by some of the warm absorber components.
    \item The X-ray broad emission features can be modeled by a third photoionized component. 
    \item  Our X-ray spectral analysis found that the line profiles of the narrow emission lines are asymmetric, the photoionized plasma is turbulent and the emission region is compact in size. Similar results have been found in the optical studies. Furthermore, distance estimates, from literature, of the X-ray and optical narrow emission line regions suggest that they might be co-located. Therefore, it is possible that the multi-phase nature of the NELR manifests its emission in different energy bands.
    \item Future missions like \textit{Arcus} are in need in order to better constraints on the distances of the warm absorber components. Hence, we can tell whether the warm absorber intervenes along the line of sight from the X-ray narrow line emitter to the observer.
    \item The X-ray NELR is not the counterpart of the UV/X-ray absorber outside the line of sight. 
    \item The \ion{N}{vii} Ly$\gamma$ to Ly$\alpha$ ratio is abnormally high in the stacked RGS spectra in 2013--2014. We investigate the possibility that this line may be produced by charge exchange.
\end{enumerate}

\begin{acknowledgements}
      This work is based on observations obtained with \textit{XMM}-Newton, an ESA science mission with instruments and contributions directly funded by ESA Member States and the USA (NASA). SRON is supported financially by NWO, the Netherlands Organization for Scientific Research. The research at the Technion is supported by the I-CORE program of the Planning and Budgeting Committee (grant number 1937/12). JM acknowledges discussions with M. Whewell. EC is partially supported by the NWO-Vidi grant number 633.042.525. SB acknowledges financial support from the Italian Space Agency under grant ASIINAFI/037/12/0, and from the European Union Seventh Framework Programme (FP7/2007-2013) under grant agreement no. 31278. LDG acknowledges support from the Swiss National Science Foundation. POP acknowledges financial support from the french CNES. GP acknowledges support from the Bundesministerium f\"{u}r Wirtschaft und Technologie/Deutsches Zentrum f\"{u}r Luft- und Raumfahrt (BMWI/DLR, FKZ 50 OR 1604) and the Max Planck Society. EB is grateful for funding from the European Union's Horizon 2020 research and innovation programme under the Marie Sklodowska-Curie grant agreement no. 655324. 
\end{acknowledgements}


\end{document}